\documentclass[a4paper,11pt]{article}
\usepackage[a4paper, margin=2cm]{geometry}
\usepackage[utf8]{inputenc}
\usepackage[T1]{fontenc}
\usepackage{amsmath}
\usepackage{amsfonts}
\usepackage{amssymb}
\usepackage{amsthm}
\usepackage{amsbsy}
\usepackage{mathrsfs}
\usepackage{slashed}
\usepackage{graphicx}
\usepackage{tikz-cd}
\usepackage{fancyhdr}
\usepackage{indentfirst} 
\usepackage{jheppub}

\newcommand{\pd}{\partial}

\newcommand{\bra}{\langle}

\newcommand{\ket}{\rangle}
\newcommand{\bx}{\square}

\newcommand{\tc}{\tilde{c}}
\newcommand{\tb}{\tilde{b}}

\numberwithin{equation}{section}

\newtheoremstyle{henrique}      
  { }                                              
  {}                                              
  {\normalfont}                            
  {}                                              
  {\bfseries}                                
  {.}                                             
  { }                                             
  {}                                              
\theoremstyle{henrique}

\hypersetup{colorlinks=false, urlbordercolor={1 0 0}, linkbordercolor={0 0 1}, citebordercolor={0 0 1}}

\title{\boldmath Field theory actions for ambitwistor string and superstring}

 \author{Nathan Berkovits}
 \author{and Matheus Lize }
 \affiliation{\textit{ICTP South American Institute for Fundamental Research,\\
Instituto de F\'{i}sica Te\'{o}rica, Universidade Estadual Paulista, \\
Rua Dr. Bento Teobaldo Ferraz 271, S\~{a}o Paulo, SP Brasil}}

\emailAdd{nberkovi@ift.unesp.br}
\emailAdd{lize@ift.unesp.br}

\abstract{We analyze the free ambitwistor string field theory action for the
bosonic string, heterotic string and both GSO sectors of the Type II string. 
The spectrum contains non-unitary states and provides an interesting consistency test for one-loop ambitwistor
string computations.}

\begin{document}
\maketitle
\flushbottom

\section{Introduction}

The ambitwistor string was introduced by Mason and Skinner in \cite{Mason:2013sva} as a string theory whose tree amplitudes reproduce the Cachazo-He-Yuan description \cite{Cachazo:2014} of massless amplitudes in ten dimensions. Like the $d=4$ twistor string of \cite{Berkovits:2004}, the $d=10$ ambitwistor string only contains left-moving variables on the worldsheet and has no massive states. Although \cite{Mason:2013sva} describes bosonic, heterotic and Type II versions of the ambitwistor string, only the GSO$(+)$ sector of the Type II version correctly describes the massless GSO(+) sector of the usual Type II superstring, i.e. $d=10$ $N=2$ supergravity. 

Nevertheless, it is interesting to identify the spectrum of massless states described by the other ambitwistor strings, i.e. the bosonic, heterotic and GSO$(-)$ sector of the Type II ambitwistor string. In this paper, these spectra will be identified using the standard BRST method where equations of motion and gauge invariances are derived from the cohomology at ghost-number 2 of the BRST operator. The quadratic kinetic term for the string field theory action will be explicitly constructed for these ambitwistor strings and expressed in a gauge-invariant manner.

Except for the GSO$(+)$ sector of the Type II ambitwistor string whose kinetic term is the usual $d=10$ $N=2$ supergravity action, the kinetic terms for the other ambitwistor strings contain higher-derivative terms which imply a non-unitary spectrum. This is similar to the $d=4$ twistor string whose spectrum includes conformal supergravity. In hindsight, this should have been expected since the three-point amplitudes in ambitwistor strings (except for the Type II string) were computed to have higher powers of momenta than the usual massless theories. And since there are no dimensionful constants like $\alpha'$ in ambitwistor strings, the higher momentum dependence in the cubic term of the string field theory action implies higher momentum dependence in the quadratic kinetic term.

Nevertheless, this non-unitary massless spectrum does not seem to have been previously analyzed for a few reasons. Firstly, vertex operators in the ambitwistor string were assumed in \cite{Mason:2013sva} to contain only $P_m$ dependence and to be independent of $\pd X^m$, where $X^m$ and $P_m$ are the spacetime variable and its conjugate momentum. Secondly, the definition of BPZ conjugate used in a previous construction of the ambitwistor string field theory action \cite{Reid-Edwards:2017goq} was chosen in an unconventional manner in order to give a kinetic term with the standard unitary massless spectrum. And thirdly, a singular gauge-fixing procedure for the ambitwistor string was adopted in \cite{Siegel:2014}\cite{Hohm:2014} which introduces non-trivial $\alpha'$ dependence into the  field theory action so that the higher-derivative cubic term does not imply a higher-derivative kinetic term. 
 
One possible application of our result is to test the consistency of one-loop amplitude prescriptions for ambitwistor string computations. Although the GSO$(+)$ sector of the Type II ambitwistor superstring is the only ambitwistor string with a conventional spectrum, one can in principle try to compute one-loop amplitudes in any of the ambitwistor strings. It would be interesting to verify if the partition functions computed using the one-loop prescriptions in \cite{ Casali:2014}\cite{Geyer:2014} reproduce the non-unitary states in the massless spectrum. Note that even for the Type II ambitwistor string,  the one-loop prescription using the RNS method involves first computing the partition function for different spin structures and then summing these partition functions. Before performing the sum over spin structures, one should be able to observe in the partition function the contribution of the states in the GSO$(-)$ sector. It would be very interesting to verify if the non-unitary spectrum of the GSO$(-)$ sector described in this paper is correctly reproduced by the one-loop computations.

A second possible application of our result is to try to generalize the quadratic kinetic term  computed here to the full string field theory action including interactions. As noted in \cite{Azevedo:2017lkz}, the $d=10$ heterotic ambitwistor string has some similarities with the $d=4 $ twistor string which describes $N=4$  $d=4 $ conformal supergravity coupled to super-Yang-Mills  \cite{Witten:2014}. It would be interesting to study if the $d=10$ heterotic ambitwistor string field theory action describes a $d=10$ generalization of $N=4$ $d=4$ conformal supergravity.

In section \ref{sectionbosonic} of this paper, we use the standard BRST method to compute the kinetic term in the bosonic ambitwistor string field theory action. And in sections \ref{sectionII} and \ref{sectionheterotic}, we repeat this procedure for the Neveu-Schwarz states in the Type II and heterotic ambitwistor string field theory actions.

\section{Bosonic ambitwistor string}
\label{sectionbosonic}
We first describe the bosonic ambitwistor string.
Subsection \ref{sectionbosonic1} defines the model and our notation, 
subsection \ref{sectionbosonic2} computes the spectrum via BRST cohomology, and subsection \ref{BA}
constructs the kinetic string field theory action.
The same steps will be later described in sections \ref{sectionII} and \ref{sectionheterotic}
for the Type II and heterotic ambitwistor strings.

\subsection{Review and notation}
\label{sectionbosonic1}
The gauge-fixed worldsheet action\cite{Mason:2013sva} is

\begin{equation}
S_{B} = \frac{1}{2\pi}\int d^{\,2}z  
	\,( P_{m} \bar{\pd}X^{m} + b\bar{\pd} c + \tilde{b}\bar{\pd}\tilde{c}  ),
\label{1}
\end{equation}

\noindent
where all matter and ghost fields are left-moving bosons and fermions on the worldsheet. $(P_{m}, X^m)$ are the matter fields of conformal weight $(1,0)$, 
$(b, c)$ are the Faddeev-Popov ghosts for reparametrization symmetry of  
conformal weight $(2,-1)$, and $(\tilde{b},\tilde{c})$ are
the Faddeev-Popov ghosts for the null geodesic constraint, $P^{2}=0$, and 
carry conformal weight $(2,-1)$. The action \eqref{1} is invariant under the BRST transformation generated by

  \begin{equation}
 Q = \oint \frac{dz}{2 \pi i}  \left( cT^{M} + cT_{\tb\tc} + bc\pd c  +\frac{1}{2}\tilde{c}P^{2} \right)
 \label{22}
 \end{equation}
 \noindent
where 

\begin{equation}
T^{M} = -P_{m} \pd X^{m}, \quad T_{\tb\tc} =  \tilde{c}\pd\tilde{b}-2\tilde{b}\pd \tilde{c},
\end{equation}
and one uses the free field OPE's,

\begin{equation}
	P_{m}(z)X^{n}(w) \sim -\frac{\delta^{n}_{\;m} }{(z-w)}, \quad b(z)c(w) \sim \frac{1}{(z-w)}, \quad \tilde{b}(z)\tilde{c}(w) \sim \frac{1}{(z-w)} .
	\label{21}
\end{equation}

 \noindent
Notice that the $XX$ OPE is regular,  so  $e^{ik\cdot X}$ does not acquire  an anomalous dimension. Furthermore, there are no dimensionful parameters such as $\alpha'$ in the theory. So the physical spectrum defined by the BRST cohomology is not expected to contain massive states. This will be confirmed below, however, we will show that the spectrum contains both unitary and non-unitary massless states.

Physical closed string states should have ghost number 2 where the ghost number is defined as 
\begin{equation}
N_{gh} = -\oint \frac{dz}{2 \pi i} (bc +\tb\tc ),
\label{18}
\end{equation} 
\noindent
such that $b,\tb$ have ghost number $-1$ and $c,\tilde{c}$ have ghost number $1$.
In order to compute the ghost number 2 cohomology, Mason and Skinner \cite{Mason:2013sva} considered only homogeneous polynomials in $P$ so that their expression for the spin-2 unintegrated vertex operator is 

\begin{equation}
V(z) = c(z)\tilde{c}(z)  P_{m}(z)P_{n}(z)g^{mn}  e^{ikX(z)}.
\label{2}
\end{equation}

\noindent
BRST closedness implies 

\begin{equation}
	k^{m}g_{mn} =0 \quad \text{ and } \quad k^{2}=0,
	\label{19}
\end{equation}

\noindent
while BRST exactness gives

\begin{equation}
	\delta g^{mn} = k^{(m}\lambda^{n)}  \quad \text{ and } \quad  k^{m} \lambda_{m} =0 .\label{25}
\end{equation}

Equations \eqref{19} and \eqref{25} are the usual conditions satisfied by  the graviton field in linearized gravity where $g_{mn}$  and $\lambda$ are  the target space metric and infinitesimal diffeomorphism generator. So it is tempting to say that the vertex \eqref{2} describes the graviton. However, this would present a paradox since the three-point scattering amplitude computed
using  \eqref{2} is
 \cite{Mason:2013sva}

\begin{equation}
\begin{aligned}
\bra V(z_{1})V(z_{2})V(z_{3})\ket 
&=\delta^{26}\left(\sum k\right)( g_{2}^{rs}  k^{1}_{r}k^{1}_{s} )(g_{3}^{mn}  k^{2}_{m}k^{2}_{n})(g_{1}^{pq}  k^{3}_{p}k^{3}_{q}).
\end{aligned}
\label{20}
\end{equation}
Since \eqref{20} behaves like $k^6$ instead of the $k^2$ behavior of general relativity and since there are no dimensionful parameters in the theory, one would expect the kinetic term for $g_{mn}$ should also behave like $k^6$.
This  suggests that the equation of motion for $g_{mn}$ should be something like $\bx^{3} g^{mn}=0$ instead of the  $\bx g_{mn}=0$ equation implied by \eqref{19}.

\noindent

In this paper, we aim to clarify this issue. Mason and Skinner constructed the vertex operator using only polynomials in $P$. However, from the string theory perspective, nothing prevents us from considering vertex operators involving $\pd X$. By considering the most general 
vertex operator with ghost number two, we will find that the equation of motion for $g_{mn}$ behaves like $k^6$.

\subsection{Bosonic   spectrum}
\label{sectionbosonic2}

The most general vertex operator with ghost number two that is annihilated by $b_{0}$ and $L_{0}$ is\footnote{Since $\bar L_0$ is identically zero, the usual constraints that $L_0 - \bar L_0$ and $b_0 - \bar b_0$ annihilate the off-shell closed string vertex operator are replaced by the constraints that $L_0$ and $b_0$ annihilate the off-shell vertex operator. 
	By $L_{0}$ and $b_{0}$ we mean the zero-modes of the $b$-ghost and stress-energy tensor:
\begin{equation}
	b_{0} = \oint \frac{dz}{2\pi i}  z  \, b(z)  \quad \text{ and } \quad L_{0} = \oint \frac{dz}{2\pi i}  z \, T(z) .
\end{equation}} 

\begin{equation}
\begin{aligned}
V(z) = & c\tilde{c}\Phi_{2}  + c\pd \tilde{c} \Psi_{1}  + \pd^{2}c\tilde{c}S^{(4)}+c\pd^{2} \tilde{c}S^{(5)}+ \pd^{2}cc  S^{(2)} + \pd \tilde{c} \tilde{c} \Gamma_{1} \\&+ \pd^{2}\tilde{c}\tilde{c}S^{(3)}+ 
  \tilde{b}\tilde{c} c\pd \tilde{c}S^{(6)} + bc \pd \tilde{c} \tilde{c}S^{(1)}\,,
\end{aligned}
\label{3}
\end{equation}

\noindent
where 

\begin{equation}
\begin{aligned}
&\Phi_{2} = P^{m}P^{n} G_{mn}^{(1)} +\pd X^{m}\pd X^{n}G^{(2)}_{mn}+\pd X^{m} P^{n}H_{mn}  +\pd^{2}X^{m}A^{(1)}_{m} +\pd P^{m}A^{(2)}_{m} ,\\
&\Psi_{1}=  P^{m}A^{(5)}_{m}+\pd X^{m}A^{(6)}_{m} \;, \;\;\;\;\; \Gamma_{1}=  P^{m}A^{(3)}_{m}+\pd X^{m}A^{(4)}_{m},\\
& H_{mn} = G^{(3)}_{mn} + B_{mn}.
\end{aligned}
\end{equation}

\noindent
The symmetric fields with two indices are represented by $G^{(1)}_{mn},G^{(2)}_{mn},G^{(3)}_{mn}$; the antisymmetric 2-form  by $B_{mn} = B_{[mn]}$;
the 1-forms by $A^{(1)}_{m},\dots ,A^{(6)}_{m}$; and the scalars by $S^{(1)},\dots ,S^{(6)}$.
These fields have arbitrary dependence on $X$, e.g., $G^{(1)}_{mn} =G^{(1)}_{mn}(X)$.

The target space fields have gauge symmetry $\delta V = Q\Lambda$,
where $\Lambda$ has ghost number one and also satisfies $b_{0}\Lambda=L_{0}\Lambda =0$.
The most general gauge parameter $\Lambda$ takes the form

\begin{equation}
\begin{aligned}
\Lambda = & c P^{m}\Lambda^{(1)}_{m} + c\pd X^{m}\Lambda^{(2)}_{m}  +\tilde{c} P^{m}\Lambda^{(4)}_{m} + \tilde{c}\pd X^{m}\Lambda^{(5)}_{m} + \pd \tilde{c} \Lambda^{(6)} +
 b c \tilde{c} \Lambda^{(7)} + c\tilde{b}\tilde{c} \Lambda^{(3)} .
 \label{4}
\end{aligned} 
\end{equation}

The vertex \eqref{3} can be simplified by removing fields that are pure gauge. 
Whenever the gauge transformation of a field does not involve spacetime derivatives of the gauge parameter,  
we can eliminate this field without producing gauge-fixing ghosts. 
By a suitable choice of gauge parameters, it is easy to show that the fields $S^{(2)},S^{(4)},S^{(6)}, A^{(1)}_{m},A^{(2)}_{m}$ can be eliminated from the vertex operator \eqref{3}.

\paragraph{Cohomology:}
Now that we have the most general vertex operator we can calculate the  cohomology. The BRST-closedness condition $QV =0$ gives the following auxiliary equations

\begin{equation}
\begin{aligned}
&  A^{(5)}_{n} = -\pd^{m}G^{(1)}_{mn} ,\quad A^{(6)}_{m}= -\frac{1}{2}\pd^{n}H_{mn},\quad A^{(3)}_{m} = A^{(6)}_{m},\quad A^{(4)}_{m} = - \pd_{m}S^{(1)},  
  \\& G^{(3)}_{mn}=\frac{1}{2}\bx G^{(1)}_{mn}  -\frac{1}{2} \pd_{(n} \pd^{r} G^{(1)}_{m)r},\qquad  2G^{(2)}_{mn}= + \frac{1}{2}\bx G^{(3)}_{mn}   + \eta_{mn}S^{(1)},
\\& S^{(5)}  =+\frac{1}{2}\pd^{n}\pd^{m}G^{(1)}_{mn},\qquad
  S^{(3)} = -\frac{1}{2}\pd^{m}A^{(3)}_{m}+\frac{3}{2} S^{(1)} , 
\label{11}
\end{aligned}
\end{equation}

\noindent
together with the equations of motion 

     \begin{equation}
     \begin{aligned}
     \bx G^{m(1)}_{m}   +4\pd^{n}\pd^{m}G^{(1)}_{mn}= & 0,\\
       \bx B_{nm} +\pd_{n}\pd^{p} B_{mp} -\pd_{m} \pd^{p}B_{np}=& 0, \\
    \bx^{3} G^{(1)}_{mn} -\bx^{2}\pd_{(n}\pd^{p}G^{(1)}_{m)p} + 4\eta_{mn}S^{(1)}+ 16\pd_{m}\pd_{n}S^{(1)} = & 0.\\
    \label{13}
     \end{aligned}
     \end{equation}
\noindent
The gauge transformations given by $\delta V = Q\Lambda$ for the propagating fields are 

  \begin{equation}
  \begin{aligned}
  &\delta G^{(1)}_{(mn)} =\frac{1}{2} \pd_{(n}\Lambda^{(1)}_{m)} -\frac{1}{6}\eta_{mn}(\pd \cdot \Lambda^{(1)}),\\
  &\delta B_{[mn]} =\pd_{[m}\Lambda^{(4)}_{n]} ,\\
  &\delta S^{(1)} = \frac{1}{24}\bx^{2} (\pd \cdot \Lambda^{(1)}).\\
  \label{12}
  \end{aligned}
  \end{equation}

\noindent
Although the gauge transformation for the field $G^{(1)}_{mn}$ does not correspond to the linear diffeomorphism of the graviton, we will perform in the next subsection a field redefinition to obtain the usual transformation. However, it is unclear how to interpret this vertex operator as a deformation around the background.

\subsection{ Ambitwistor  kinetic term }
\label{BA}

The  standard  kinetic term $S[\Psi] =   \frac{1}{2}\bra \Psi | (c_{0}- \bar c_0) Q\Psi \ket$ for the closed bosonic string was introduced in \cite{Zwiebach:1992ie} using the string field defined by the state-operator mapping: $ |\Psi \ket = V(0)|0\ket$ where $|0\ket$ is the $SL(2,C)$ vacuum and $|\Psi\ket $ is constrained to satisfy
$( L_{0} - \bar{L}_{0})|\Psi\ket =( b_{0} - \bar{b}_{0})|\Psi\ket =0$.
For the ambitwistor string, we will have a similar kinetic term; however, since all the fields are holomorphic,
we discard the antiholomorphic zero-modes $\bar{L}_{0}$ and $\bar{b}_{0}$. 

Therefore, we propose for the ambitwistor string kinetic term 

\begin{equation}
S[\Psi] =   \frac{1}{2}\bra \Psi | c_{0} Q\Psi \ket    = \frac{1}{2} \bra I \circ V(0)|  \pd cQV(0) \ket 
\label{8}
\end{equation}

\noindent
where $|\Psi\ket $ is constrained to satisfy

\begin{equation}
L_{0} |\Psi \ket = b_{0} |\Psi \ket=0.
\label{5}
\end{equation}

\noindent
The bra state of the string field $\bra \Psi |$ is defined by the usual BPZ conjugate, $\bra \Psi | =\bra 0 | I \circ V(0)$ where $I(z) = 1/z$. 
For a primary field of conformal weight $h$ the conformal transformation $I$ acts as
\begin{equation}
I \circ \phi(y) = (\pd_{y} I)^{h}\phi(1/y).
\label{24}
\end{equation}

The variation of $S[\Psi]$ implies $c_{0}Q |\Psi \ket =0$. The condition $b_{0}|\Psi\ket =0$
turns this into the linearized equations of motion $Q|\Psi \ket =0$.
The action $S[\Psi]$ is invariant under $|\delta\Psi\ket =Q|\Lambda\ket$, where $\Lambda$ has ghost number one and is annihilated by $L_{0}$ and $b_{0}$.
The proof of gauge invariance and the derivation of the field equations follows exactly as in \cite{Zwiebach:1992ie}, so it will not be reproduced here.
A similar string field theory action was previously proposed in \cite{Reid-Edwards:2017goq}, but their construction did not allow insertions of $\pd X$ in the vertex operator and
they modified the usual definition of the BPZ inner product to get a massless unitary spectrum.

Let us focus on computing the action for the ambitwistor string vertex operator \eqref{3}.
The action can be calculated in two different -- but equivalent -- ways: using creation and annihilation operator algebra or vertex correlation functions.
We will work with the latter.

The gauge parameter  \eqref{4} can set $S^{(2)},S^{(4)},S^{(6)}, A^{(1)},A^{(2)}$ to zero without producing ghosts, so the vertex operator \eqref{3} simplifies to
 
\begin{equation}
\begin{aligned}
V(z) = & c\tilde{c}\Phi_{2}  + c\pd \tilde{c} \Psi_{1}  +c\pd^{2} \tilde{c}S^{(5)} + \pd \tilde{c} \tilde{c} \Gamma_{1} + \pd^{2}\tilde{c}\tilde{c}S^{(3)}+ bc \pd \tilde{c} \tilde{c}S^{(1)}\,,
\end{aligned}\label{6}
\end{equation}

\noindent
where 

\begin{equation}
\begin{aligned}
&\Phi_{2} = P^{m}P^{n} G_{mn}^{(1)} +\pd X^{m}\pd X^{n}G^{(2)}_{mn}+\pd X^{m} P^{n}H_{mn} ,\\  &\Psi_{1}=  P^{m}A^{(5)}_{m}+\pd X^{m}A^{(6)}_{m},\;\;\;\Gamma_{1}=  P^{m}A^{(3)}_{m}+\pd X^{m}A^{(4)}_{m}.\\
\end{aligned}
\end{equation}

One can verify that the auxiliary field equations of \eqref{11} imply that
\begin{equation}
\begin{aligned}
T(z)V(0) \sim  
& +z^{-4}[  -c\tilde{c}(H_{m}^{m} + 6 S^{(5)} ) ]\quad +\quad z^{-3}[c\pd \tilde{c}( -\pd^{m} A^{5}_{m} -2 S^{(5)}    ) ]\;\;+ \\
&+ z^{-3}[c\tilde{c}(-2P^{m}(\pd^{n}G^{1}_{mn}+A_{m}^{(5)}) -\pd X^{m}(\pd^{n}H_{mn}+2A^{(6)}_{m}) )   ]\;\;+\\
 &+  
 z^{-3}[\tilde{c}\pd \tilde{c}( +\pd^{m} A^{(3)}_{m}  +2 S^{(3)} -3S^{(1)} ) ] + z^{-1}\pd V(0) \\
\sim& z^{-4}[  -c\tilde{c}(H_{m}^{m} + 6 S^{(5)} ) ]\;\; +z^{-1}\pd V(0).\\
\end{aligned}
\end{equation}

\noindent
So after applying the auxiliary field equations of  \eqref{11}, $T$ has no double or cubic poles with $V$, which implies that $I\circ V(z) = V(I(z))$ and the string action \eqref{8} becomes  the two point function $\bra V(I(0)) \pd c QV(0)\ket$.
We stress that applying the auxiliary field equations before computing the kinetic term is a trick to simplify the computation.
One could have done the calculation in full detail and obtained the same answer.

Using the vacuum normalization $\bra \pd^{2}c\pd c c \pd^{2}\tc\pd \tc \tc\ket= 4$, the string action becomes

    \begin{equation}
    \begin{aligned}
            S =  - \int d^{26}X  
             &\left[ +\frac{1}{8}  G^{mn(1)} \bx^{3}G^{(1)}_{mn} + \frac{1}{4} \pd_{r} G^{mr(1)} \bx^{2}\pd^{p}G^{(1)}_{mp}  +4G^{mn(1)}\pd_{n}\pd_{m}S^{(1)}  +\right.\\
       &\left.+G^{p(1)}_{p}\bx S^{(1)} -\frac{1}{2} B^{mn}(\bx B_{mn} +\pd_{[m} \pd^{p}B_{n]p}) \right].
    \end{aligned}
    \label{7}
     \end{equation}

 \noindent 
The equations of motion agree with \eqref{13} and the gauge transformations are those given by \eqref{12}. Note that the kinetic action for $G^{(1)}_{mn}$ involves 6 derivatives, so the inconsistency between the momentum dependence of the 3-point amplitude \eqref{20} and the momentum dependence of the kinetic term is resolved. 
 
To write the kinetic action in terms of gauge invariant objects, it is convenient to perform a field redefinition since
 the gauge transformation for $G_{mn}^{(1)}$ is not quite the transformation of the graviton. 
A convenient field redefinition is
 
 \begin{equation}
 h_{mn} - \frac{1}{6}\eta_{mn} h_p^p= G^{(1)}_{mn}  ,\quad  t = 4S^{(1)} - \frac{1}{6}\bx^{2} h_p^p, 
 \end{equation}

 \noindent
 to obtain the gauge transformations of linearized gravity
 
 \begin{equation}
 \begin{aligned}
 &\delta h_{mn} = \frac{1}{2} \pd_{(n}\lambda_{m)} \;, \;\;\; \delta t = 0.
 \end{aligned}
 \end{equation}
 
\noindent 
 The action \eqref{7} written in terms of gauge invariant objects becomes 
 
 \begin{equation}
 S= -\int d^{26}X \left[ \frac{1}{2}R_{mn}\bx R^{mn} - \frac{1}{4}R\bx R + tR  - \frac{1}{3!} H^{mnp}H_{mnp}\right],
 \end{equation}

\noindent 
where we have defined the linearized  Ricci tensor and 3-form field strength 

  \begin{equation}
 \begin{aligned}
 & 2R_{mn} = \pd_{m} \pd^{p}h_{np}+\pd_{n} \pd^{p}h_{mp} - \bx h_{mn} - \pd_{m}\pd_{n} h_p^p,\\
 & H_{mnp} = \pd_{m} B_{np} +\pd_{n} B_{pm} + \pd_{p} B_{mn}  .
 \end{aligned}
 \end{equation}

\noindent
One can simplify further  by shifting  $ t$ to $ t + \bx R/4$ so the term $R\bx R$ drops out of the action.

\section{ Type II ambitwistor  }
\label{sectionII}

In this section we will describe the Type II ambitwistor string for both $GSO$ Neveu-Schwarz sectors. The spectrum for the $GSO(+)$ Neveu-Schwarz sector will be the usual bosonic massless Type II supergravity states, however, the spectrum for the $GSO(-)$ Neveu-Schwarz sector will have some unusual non-unitary states.
Although only the $GSO(+)$ sector is supersymmetric, the $GSO(-)$ sector is expected to appear as intermediate states before summing over spin structures using the RNS formalism. So by analyzing the contribution of individual spin structures to the one-loop partition function of the Type II ambitwistor superstring, one should be able to verify this unusual spectrum for the $GSO(-)$ sector.

\subsection{Review and notation }

For the Type II action we add two fermionic holomorphic worldsheet variables $\psi_{1},\psi_{2}$, both with conformal weight $1/2$. We also introduce two pairs of bosonic Faddev-Popov ghosts: $(\beta_{1},\gamma_{1})$ and $(\beta_{2},\gamma_{2})$.  The $\beta$'s have conformal weight $3/2$ while the $\gamma$'s have conformal weight $-1/2$.
The action for this system is

\begin{equation}
S_{tII} = \frac{1}{2\pi}\int d^{\,2}z  \,( P_{m} \bar{\pd}X^{m} + b\bar{\pd} c + \tilde{b}\bar{\pd}\tilde{c} + \psi_{1}\bar{\pd}\psi_{1}+\psi_{2}\bar{\pd}\psi_{2} +\beta_{1}\bar{\pd}\gamma_{1} + \beta_{2}\bar{\pd}\gamma_{2}).
\label{atII}
\end{equation}

\noindent
The new field variables have the OPE's

\begin{equation}
\psi^{m}_{i}(z)\psi^{n}_{j}(w) \sim \delta_{ij}\frac{\eta^{mn}}{(z-w)}\;, \quad \beta_{i}(z)\gamma_{j}(w) \sim -\frac{\delta_{ij}}{(z-w)} \quad \text{for} \quad i,j=1,2,
\end{equation}

\noindent 
in  addition to the ones obtained in \eqref{21}.
The action \eqref{atII} also  presents BRST symmetry generated by

\begin{equation}
Q = \oint \frac{dz}{2\pi i} (cT^{M} + cT_{\tb\tc} + cT_{\beta_{1}\gamma_{1}}+cT_{\beta_{2}\gamma_{2}}+ bc\pd c + \frac{1}{2}\tilde{c}P^{2} + \gamma_{1} P\cdot\psi_{1} + \gamma_{2}P\cdot\psi_{2} -\gamma_{1}^{2}\tilde{b}- \gamma_{2}^{2}\tilde{b} ),
\label{23}
\end{equation}  

\noindent
where 

\begin{equation}
\begin{aligned}
& T^{M} = -P_{m} \pd X^{m} -\frac{1}{2}\psi_{1}\cdot\pd\psi_{1}-\frac{1}{2}\psi_{2}\cdot\pd\psi_{2}  ,\quad T_{\tb\tc} =  \tilde{c}\pd\tilde{b}-2\tilde{b}\pd \tilde{c} ,\\ & T_{\beta_{i}\gamma_{i}}= -\frac{1}{2}\pd \beta_{i}\gamma_{i} - \frac{3}{2}\beta_{i}\pd \gamma_{i} \;.
\label{27}
\end{aligned}
\end{equation}

\noindent
The nilpotency of  the BRST charge imposes  the critical  spacetime  dimension  $d=10$. In order  to write the vertex operator in the picture $(-1,-1)$ we  bosonize the ghosts $(\beta_{i},\gamma_{i})$ by introducing  a set of fermions $(\eta_{i},\xi_{i})$ with conformal weight $(1,0)$ together with a chiral boson $\phi_i$. This
new system is described by the free field OPE's

\begin{equation}
\phi_{i}(z)\phi_{j}(w) \sim - \delta_{ij}\ln(z-w),
\quad 
\eta_{i}(z)\xi_{j}(w) \sim \frac{\delta_{ij}}{z-w},
\end{equation} 

\noindent
and the change of variables is

\begin{equation}
\beta_{i} = e^{-\phi_{i}}\pd \xi_{i} \;,\quad\gamma_{i} = \eta_{i}e^{+\phi_{i}}.
\label{14}
\end{equation}

\noindent
The BRST charge \eqref{23} in terms of bosonized variables  $(\eta,\xi,\phi)$ is  written by replacing

\begin{equation}
T_{\beta_{i}\gamma_{i}} = -\frac{1}{2}\pd\phi_{i}\pd\phi_{i} -\pd^{2}\phi_{i} - \eta_{i}\pd \xi_{i}  \quad \text{and} \quad  \gamma_{i} ^{2} = \eta_{i}\pd \eta_{i} e^{-2\phi_{i}},
\end{equation}

\noindent
for each pair $(\beta_{i},\gamma_{i})$. The ghost number charge \eqref{17} is modified to accommodate the $(\beta,\gamma)$ system as

\begin{equation}
N_{gh} = -\oint \frac{dz}{2\pi i}(bc +\tb\tc + \xi_{1} \eta_{1}+\xi_{2} \eta_{2})
\end{equation}

\noindent
 In addition to the ghost number charge we define the picture number:
\begin{equation}
 N_{P_{i}} = \oint \frac{dz}{2\pi i} (\xi_{i} \eta_{i} - \pd \phi_{i}),
\label{28}
\end{equation}
\noindent
 such that $\beta$ and $\gamma$ have picture zero and ghost number $-1$ and $1$ respectively.

\subsection{Type II  spectrum}

 There are two sectors for Neveu-Schwarz states in superstring theory which contain either GSO parity $+$ or GSO parity $-$.
  The vertex operator considered by Mason and Skinner \cite{Mason:2013sva} is in the $GSO(+)$ sector. The field content found in \cite{Mason:2013sva} is  a spin-2 $G_{mn}$,  a scalar $G^{m}_{m}$ and a 2-form $B_{mn}$ which agrees with the bosonic fields of $d=10$ N=2 supergravity. However, the ambitwistor superstring also has a $GSO(-)$ sector that has not yet been fully investigated.

 In order to distinguish the two sectors, we introduce the  operator $(-)^{parity}$ where the parity of $\psi_1$ and $e^{\phi_1}$ is defined to be odd,  the parity of $\psi_2$ and $e^{\phi_2}$ is defined to be even, and the parity of all other variables $(P_m, X^m, b, c, \tilde b, \tilde c, \xi_i, \eta_i)$ is defined to be even. One can easily verify that $(-)^{parity}$ commutes with the BRST charge of \eqref{23}.

Although the superstring is only spacetime supersymmetric after truncating out the $GSO(-)$ sector, it will be interesting to compute the spectrum for both sectors. 
The most general Neveu-Schwarz vertex operator  in the picture $(-1,-1)$ with ghost number two and which is annihilated by $b_0$ and $L_0$ is

\begin{equation}
\begin{aligned}
V(z) = & e^{-\phi_{1}}e^{-\phi_{2}} (c\tilde{c} \Phi_{1} + c \pd\tilde{c} S^{(1)}  +\tilde{c}\pd\tilde{c} S^{(6)} ) + \pd\phi_{1}e^{-\phi_{1}}e^{-\phi_{2}} c\tilde{c}S^{(2)} +\\+& e^{-\phi_{1}}\pd\phi_{2}e^{-\phi_{2}} c\tilde{c}S^{(3)} +\pd\xi_{1}e^{-2\phi_{1}}e^{-\phi_{2}}(c\tilde{c}\pd \tilde{c}\psi_{1}\cdot A^{(3)}+c\tilde{c}\pd \tilde{c}\psi_{2}\cdot A^{(4)} ) +\\ 
+& e^{-\phi_{1}}\pd\xi_{2}e^{-2\phi_{2}}(c\tilde{c}\pd \tilde{c}\psi_{1}\cdot A^{(5)}+c\tilde{c}\pd \tilde{c}\psi_{2}\cdot A^{(6)})+ \eta_{1}\pd\xi_{2}e^{-2\phi_{2}} c\tilde{c} S^{(4)} +
\\+& \pd\xi_{1}e^{-2\phi_{1}}\eta_{2} c\tilde{c} S^{(5)}, 
\label{VII}
\end{aligned}
\end{equation}
with 
\begin{equation}
\begin{aligned}
&\Phi_{1} = P\cdot A^{(1)}+\pd X\cdot A^{(2)} + B_{mn}^{(1)}\psi_{1}^{m}\psi_{1}^{n} +B_{mn}^{(2)}\psi_{2}^{m}\psi_{2}^{n}  +H_{mn}\psi_{1}^{m}\psi_{2}^{n},
\\& H_{mn} = G_{mn} + B_{mn}.
\end{aligned}
\end{equation}
\noindent
where the fields are represented by six scalars $S$,  six 1-forms $A_{m}$, one  symmetric two-form $G_{mn}$,  and three antisymmetric 2-forms  $B_{mn}$. 
Note that the vertex operator \eqref{VII} is defined in the small Hilbert space, i.e does not contain the zero mode of $\xi_i$.
Using the definition of the operator  $(-)^{parity}$ the fields can be separated into

\begin{equation}
\begin{aligned}
& GSO(+) :  H_{mn}= G_{mn} + B_{mn}, A^{(4)}_{m},A^{(5)}_{m},S^{(4)},S^{(5)}\\
& GSO(-) : A^{(1)}_{m},A^{(2)}_{m},A^{(3)}_{m},A^{(6)}_{m},B^{(1)}_{mn},B^{(2)}_{mn},S^{(1)},S^{(2)},S^{(3)},S^{(6)}.
\end{aligned}
\end{equation}

\paragraph{Cohomology:}

 As in the bosonic case, the fields in \eqref{VII} have gauge transformations $\delta V = Q\Lambda$, where the gauge field $\Lambda$ is in the small Hilbert space and satisfies $L_{0} \Lambda=b_{0} \Lambda=0$. So the gauge field with ghost number one is

\begin{equation}
\begin{aligned}
\Lambda &=  \pd \xi_{1} e^{-2\phi_{1}}e^{-\phi_{2}}c\tilde{c}( \psi_{1}\cdot \Lambda^{(1)} + \psi_{2}\cdot \Lambda^{(2)} )+\pd \xi_{2} e^{-2\phi_{2}}e^{-\phi_{1}}c\tilde{c}( \psi_{1}\cdot \Lambda^{(3)} + \psi_{2}\cdot \Lambda^{(4)} ) +\\&+ e^{-\phi_{1}}e^{-\phi_{2}}(c\Lambda^{(6)}+ \tilde{c}\Lambda^{(7)}) 
 +\pd \xi_{1}e^{-2\phi_{1}}\pd \xi_{2}e^{-2\phi_{2}} c\tilde{c}\pd \tilde{c} \Lambda^{(5)}
+\\&+\pd^{2} \xi_{1}\pd \xi_{1}e^{-3\phi_{1}}e^{-\phi_{2}} c\tilde{c}\pd \tilde{c}\Lambda^{(8)}
+\pd^{2} \xi_{2}\pd \xi_{2}e^{-3\phi_{2}}e^{-\phi_{1}} c\tilde{c}\pd \tilde{c}\Lambda^{(9)},
\end{aligned}
\end{equation}

\noindent 
which can be used to gauge away $(A^{(1)}_{m},S^{(1)},S^{(2)},S^{(5)})$. 
After using $QV=0$ to eliminate the auxiliary fields in the vertex operator \eqref{VII} whose equations of motion do not involve derivatives,  the remaining equations of motion and gauge transformations  for  both sectors are

 \begin{itemize}
   \item   $ GSO(+): $
   \end{itemize}

\begin{align*}
 \textbf{Field equations}& & &\textbf{Gauge transformations}\\
 \bx G_{mn} -\pd_{(m}\pd^{p}G_{n)p} +\pd_{n}\pd_{m}S^{(4)}&=0, &  &\delta G_{mn} =+ \frac{1}{2}\pd_{(m}\Lambda_{n)}^{(2)}  + \frac{1}{2}\pd_{(m}\Lambda_{n)}^{(3)},  \\
\pd^{p}\pd^{m}G_{pm} -\bx S^{(4)}&=0,& &\delta B_{mn} =+ \frac{1}{2}\pd_{[m}\Lambda_{n]}^{(2)}  -\frac{1}{2} \pd_{[m}\Lambda_{n]}^{(3)},\\
\bx B_{mn} + \pd_{[m}\pd^{p}B_{n]p} &=0,& &\delta S^{(4)} = \pd \cdot \Lambda^{(3)}+\pd \cdot \Lambda^{(2)}.\\ \\
\end{align*}

 \begin{itemize}
   \item   $ GSO(-): $
   \end{itemize}

    \begin{align*}
     \textbf{Field equations}& & &\textbf{Gauge transformations}\\
  \pd^{p}B_{pn}^{+}  =0,& & &\delta B^{+}_{mn} = 0, \\
     \bx B_{mn}^{+} + \pd_{[n}A^{(2)}_{m]}=0, & & &\delta B^{-}_{mn} = \pd_{[n}\Lambda_{m]}^{(4)},\\
    \bx B_{mn}^{-} - \pd_{[n}\pd^{p}B^{-}_{m]p}  =0, & & &\delta A^{(2)}_{m} = -4\pd_{m}\Lambda^{(9)} -\pd_{m}\pd \cdot \Lambda^{(4)}, \\
    \end{align*}

\noindent
where in the $GSO(-)$ sector we defined $B_{mn}^{\pm} \equiv B_{mn}^{(1)} \pm B_{mn}^{(2)}$. The field content in the $GSO(+)$ sector is the expected one from superstring theory and has a graviton $G_{mn}$ coupled to a scalar $S^{(4)}$, and an antisymmetric 2-form $B_{mn}$. On the other hand, the spectrum in the $GSO(-)$ sector  is unusual and includes two antisymmetric 2-forms and a 1-form. One of the antisymmetric 2-forms has the usual gauge transformation but the other one is gauge invariant.

\subsection{Ambitwistor kinetic term  }
\label{TIIkinect}

The construction of the quadratic action for the superstring is similar to the bosonic construction of section \ref{BA}.
 In addition to the constraints $L_{0}| \Psi \ket =b_{0}|\Psi \ket=0 $, the string field at ghost-number 2
  is also constrained to be in the $(-1,-1)$ picture in the small Hilbert space. 
 The string field  $|\Psi\ket$ is given by the vertex operator \eqref{VII} introduced in the previous section. We have

\begin{equation}
S[\Psi] = \frac{1}{2}\bra \Psi |  c_{0} Q |\Psi  \ket =\frac{1}{2}\bra I \circ V(0) | \pd c Q V(0)  \ket  
\label{16}
\end{equation}
\noindent
where $I\circ V(z)$ is the conformal transformation \eqref{24}. The vertex operator \eqref{VII}, after eliminating gauge fields and auxiliary fields, is a primary field with conformal weight zero, i.e, 

\[T(z)V(0) \sim z^{-1}\pd V(0),\]
\noindent
thus the conformal transformation $I\circ V(z)=V(z^{-1}) $  acts as \eqref{24}. So the calculation for the action becomes an ordinary two point function  with vacuum normalization $\bra c\pd c \pd^{2} c \tilde{c}\pd \tilde{c}\pd^{2} \tilde{c} e^{-2\phi_{1}}e^{-2\phi_{2}}\ket  = 4$. After some algebra, the actions for the $GSO(\pm)$ Neveu-Schwarz sectors are

\begin{equation}
\begin{aligned}
S^{+}= -\frac{1}{2}\int d^{10}x & \left[  G^{mn}(\frac{1}{2} \bx G_{mn} -\frac{1}{2} \pd_{(m}\pd^{p}G_{n)p} ) +S^{(4)}(\pd^{p}\pd^{m}G_{pm} -\frac{1}{2}\bx S^{(4)})+\right.\\
& \left.\quad\qquad\qquad\qquad\qquad\qquad\qquad +B^{mn}(\frac{1}{2} \bx B_{mn} 
  +\frac{1}{2} \pd_{[m}\pd^{p}B_{n]p})\right],
\label{26}
\end{aligned}
\end{equation}

\begin{equation}
\begin{aligned}
S^{-}= -\frac{1}{2}\int d^{10}x & \left[B^{mn(1)}( \bx B_{mn}^{(1)} - \pd_{[n}\pd^{p}B^{(1)}_{m]p}+\pd_{[n}A^{(2)}_{m]})+\right.
\\ &\left.\qquad\qquad\qquad\qquad +B^{mn(2)}( \bx B_{mn}^{(2)} - \pd_{[n}\pd^{p}B^{(2)}_{m]p}+ \pd_{[n}A^{(2)}_{m]})  \right].
\label{29}
\end{aligned}
\end{equation}

The $GSO(+)$ sector has the standard Type II  spectrum  -- graviton, Kalb-Ramond, and dilaton.  In order to make the field content more clear, rewrite the action \eqref{26} in terms of gauge invariant objects by redefining the fields
\noindent
\begin{equation}
\begin{aligned}
& G_{mn} = h_{mn}, \quad   R = -\bx h^p_p + \pd^{m}\pd^{n}h_{mn}, \quad\phi = S^{(4)} + h_{m}^{m}, \quad \\& H_{mnp} = \pd_{m}B_{np} +\pd_{n}B_{pm} +\pd_{p}B_{mn},
\end{aligned}
\end{equation}
such that the gauge transformations are 

\begin{equation}
\delta h_{mn} =+ \frac{1}{2}\pd_{(m}\lambda_{n)}\;\;\; , \;\;\; \delta B_{mn} =+ \frac{1}{2}\pd_{[m}\omega_{n]}\;\;\; , \;\;\; \delta \phi = 0,
\end{equation}
with $\lambda_{m} = \Lambda_{m}^{(2)} + \Lambda_{m}^{(3)}$ and $\omega_{m} = \Lambda_{m}^{(2)} - \Lambda_{m}^{(3)}$. The action for the $GSO(+)$ sector written in term of these gauge covariant objects is

\noindent
\begin{equation}
\begin{aligned}
S^{+}= -\frac{1}{2}\int d^{10}x & \left[ h^{mn}\frac{1}{2} \bx h_{mn} + (\pd^{p}h_{np} )^{2} -\frac{1}{2}h_r^r\bx h_p^p+h_r^r \pd^{p}\pd^{m}h_{pm}+\phi R \right.
\\&\left.\qquad\qquad\qquad\qquad\qquad\qquad\qquad\qquad-\frac{1}{2}\phi\bx \phi+  \frac{1}{6}H^{mnp}H_{mnp}\right]
\end{aligned}
\end{equation}
which  agrees with the action found by\cite{Reid-Edwards:2017goq}. 

On the other hand, the action \eqref{29} for the $GSO(-)$ sector is unusual. In terms of $B^{\pm}_{mn} =B^{(1)}_{mn} \pm B^{(2)}_{mn} $, the action \eqref{29} is 
\begin{equation}
S^{-}= -\frac{1}{2}\int d^{10}x [\frac{1}{3!}H^{-\,mnp}H^{-}_{mnp}+\frac{1}{3!}H^{+\,mnp}H^{+}_{mnp}+B^{+\,mn}  F_{mn} ]
\end{equation}
\noindent
where $F_{mn} = \pd_{[m}A^{(2)}_{n]}$ and $H^{\pm}_{mnp}= \pd_{[m} B^{\pm}_{np]}$. So $B^-_{mn}$ has the standard kinetic term for an antisymmetric two-form, but $B^+_{mn}$ couples to $F^{mn}$ and does not have the usual gauge invariance of an antisymmetric two-form.

\section{Heterotic ambitwistor string}
\label{sectionheterotic}

\subsection{Review and notation}
The worldsheet action for the heterotic model is similar to the Type II, but the two worldsheet fermions $(\psi_{1},\psi_{2})$ are replaced by one worldsheet fermion $\psi$ together with a new  current action $S_{J}$

\begin{equation}
S_{het} = \frac{1}{2\pi}\int d^{\,2}z  \,( P_{m} \bar{\pd}X^{m} + b\bar{\pd} c + \tilde{b}\bar{\pd}\tilde{c} + \psi\bar{\pd}\psi +\beta\bar{\pd}\gamma) + S_{J}.
\label{ahet}
\end{equation}
The particular form of the current action  $S_{J}$ is irrelevant, except that it should allow the vertex operator to be written using a current  algebra $J^{a}$ which has conformal weight one and satisfies the OPE

\begin{equation}
J^{a}(z)J^{b}(w) \sim \frac{\delta^{ab}}{(z-w)^{2}} + \frac{f^{ab}_{c}}{z-w}J^{c}(w),
\end{equation} 
\noindent
where $f^{ab}_c$ are the structure constants of the algebra.
 The action \eqref{ahet} has BRST symmetry generated by 

\begin{equation}
Q = \oint dz (cT^{M} +  bc \pd c + cT_{\tilde{b}\tilde{c}} + cT_{\beta \gamma} + cT_{J}+ \frac{1}{2}\tilde{c}P^{2} + \gamma P\cdot\psi  -\gamma^{2}\tilde{b} ),
\end{equation}

\noindent
with 
\[T^{M} = -P \cdot \pd X -\frac{1}{2}\psi\cdot\pd\psi,\quad
T_{\tilde{b}\tilde{c}}= \tilde c \pd \tilde b -2 \tilde b\pd \tilde c, \quad
T_{\beta \gamma} = -\frac{1}{2} \pd\beta \gamma -\frac{3}{2} \beta\pd \gamma
,\]
\noindent
 being the stress energy tensor for the matter and ghost fields. The new feature compared to the Type II ambitwistor, after removing the variables $(\psi_{2},\gamma_{2},\beta_{2})$, is the stress energy tensor $T_{J}$ associated with the current action $S_{J}$ with
 
  \[ T_{J}(z)T_{J}(w)\sim \frac{c_{J}}{2(z-w)^{4}}+ \frac{2T_{J}(w)}{(z-w)^{2}} +\frac{\pd T_{J}(w)}{(z-w)} ,   \] 
  \noindent
  where $c_{J}$ is the central charge. Nilpotency of the BRST charge implies $41 - c_{J} -\frac{5}{2}D =0$, so the critical spacetime dimension is $D= 10$ for $c_{J} = 16$.

\subsection{Heterotic spectrum}

Although the Yang-Mills vertex operator of \cite{Mason:2013sva} for the heterotic ambitwistor string has the expected behavior for Yang-Mills scattering amplitudes, the graviton vertex operator proposed by Mason and Skinner \eqref{1} for the heterotic model has  similar issues as in the bosonic model. The three-point graviton scattering amplitude behaves like $k^4$ as opposed to the expected $k^2$ behavior of general relativity. 
After allowing  $\pd X$ in the construction of the vertex operator, we will find that the equation of motion for the symmetric 2-form $h_{mn}$  is \[\bx^{2}h_{mn} + \cdots =0,\] 
which is consistent with the momentum behavior of the three-point amplitude. Another unexpected feature of the heterotic ambitwistor string is that the spectrum contains a three-form which is not present in the massless sector of the usual heterotic superstring.

The most general vertex operator in picture $(-1)$ in the small Hilbert space that is annihilated by $b_{0}$ and $L_{0}$ with ghost number 2 is: 

\begin{equation}
\begin{aligned}
V(z) = & e^{-\phi} (c\tilde{c} \Phi_{3/2} +  c \pd\tilde{c} A^{(2)}\cdot\psi  + \pd \tilde{c}\tilde{c} A^{(1)}\cdot\psi) +\pd\phi e^{-\phi}(c\tilde{c}A^{(3)}\cdot \psi)+ \\&+
\pd \xi e^{-2\phi}(\pd \tilde{c}\tilde{c} c\Psi_{1}  + \pd^{2}\tilde{c}\tilde{c}cS^{(4)} ) + \eta(cS^{(1)}+ \tilde{c}S^{(3)})+\pd \xi e^{-2\phi}( \pd^{2}cc\tilde{c}S^{(2)} ) +\\&+ \pd\xi \pd\phi e^{-2\phi}\pd \tilde{c}  \tilde{c}c S^{(5)}+
\pd^{2}\xi e^{-2\phi}( \pd \tilde{c}\tilde{c}cS^{(6)}),
\label{het}
\end{aligned}
\end{equation}
where 
\begin{equation}
\begin{aligned}
\Phi_{3/2} &= H^{(1)}_{mn}P^{m}\psi^{n} +H^{(2)}_{mn}\pd X^{m}\psi^{n}+ C_{mnp} \psi^{m}\psi^{n}\psi^{p}  +J^{a}\psi \cdot A^{a} +  \pd\psi\cdot A^{(4)} , \\
 \Psi_{1} &= P\cdot A^{(5)}+\pd X\cdot A^{(6)} + J^{a}C^{a}+B_{mn}^{(3)}\psi^{m}\psi^{n},\quad H^{(i)}_{mn} = G^{(i)}_{mn} + B^{(i)}_{mn}.\\
\end{aligned}
\end{equation}
The target space fields are described by six abelian scalars $S$, one non-abelian scalar $C^a$, six abelian 1-forms $A_{m}$, one non-abelian 1-form $A_m^a$,  two symmetric 2-forms $G_{mn}$, three antisymmetric 2-forms $B_{mn}$ and a  3-form $C_{mnp}$.

\paragraph{Cohomology:}

The gauge invariance $\delta V= Q\Lambda$ can be used to gauge away $S^{(2)},S^{(1)},A_{m}^{(4)}$, $A_{m}^{(3)}, B^{(1)}_{mn}$ where the gauge parameter in picture $(-1)$ with ghost number 1 is

\begin{equation}
\begin{aligned}
\Lambda &= e^{-\phi} (c \Lambda^{(6)}_{m}\psi^{m}  + \tilde{c}\Lambda^{(7)}_{m}\psi^{m}) +  \pd \xi e^{-2\phi} (c\tilde{c} \Phi_{1}  + c\pd \tilde{c}\Lambda^{(2)}) +\pd^{2} \xi e^{-2\phi}c \tilde{c}\Lambda^{(8)}+\\
& +\pd^{2}\xi\pd \xi e^{-3\phi}\pd \tilde{c} \tilde{c}c \Lambda^{(10)}_{m} \psi^{m} + \pd\xi \pd\phi e^{-2\phi}c\tilde{c}\Lambda^{(9)},  \\ 
\end{aligned}
\end{equation}
\noindent
with $\Phi_{1} = P\cdot\Lambda^{(3)}+ \pd X\cdot\Lambda^{(4)} +\psi^{m}\psi^{n}\Lambda^{(5)}_{mn} + J^{a}\Lambda^{a(1)}$.

After using $QV=0$ to fix all auxiliary fields whose equations do not contain derivatives, the remaining dynamical fields are  $G^{(1)}_{mn}, G^{r(2)}_{r}, B^{(2)}_{mn}, A^{a}_{m}$  and $C_{mnp}$. The equations of motions  together with its gauge transformations for these remaining fields are

\noindent
\begin{equation}
\begin{aligned}
-\frac{1}{4}\bx^{2} G^{(1)}_{mn}   +\bx\frac{1}{4}\pd_{(m} \pd^{p}G^{(1)}_{n)p}    -\frac{1}{10}\eta_{mn}\bx\pd^{r}\pd^{s}G^{(1)}_{rs} - \frac{1}{5} \pd_{n}\pd_{m} \pd^{r}\pd^{s}G_{rs}^{(1)}+\\-\frac{1}{20}\eta_{mn}\bx G^{r(2)}_{r}  -\frac{1}{10} \pd_{n}\pd_{m}G^{r(2)}_{r}=& 0, 
\end{aligned}
\label{eom}
\end{equation}

\begin{equation}
\begin{aligned}
\bx A^{a}_{m} -\pd_{m}(\pd^{p}A^{a}_{p}) =& 0, \\
-\bx C_{mnp} +\frac{1}{6}\pd_{[p}B^{(2)}_{mn]} =& 0, \\
  \pd^{p}C_{mnp}  =& 0, \\
\end{aligned}
\label{eom2}
\end{equation}
with gauge transformations

\begin{equation}
\begin{aligned}
\delta G^{(1)}_{(mn)} &= -\frac{1}{2}\pd_{(n}\Lambda_{m)} + \frac{1}{4}\eta_{mn}\pd \cdot \Lambda , \quad \Lambda_{m} = \Lambda^{(6)}_{m}+\Lambda^{(3)}_{m},\\
\delta G^{m(2)}_{m}&=+\frac{1}{4}\bx\pd\cdot\Lambda,\\
\delta B^{(2)}_{mn} &= \pd_{[m}\Lambda^{(4)}_{n]},\\
\delta C_{mnp} &= 0,\\
\delta A^{a}_{m} &= -\pd_{m}\Lambda^{a(1)}.
\end{aligned}
\end{equation}

\subsection{Ambitwistor kinetic term }

The kinetic term follows exactly the Type II construction of section \ref{TIIkinect}, so we shall not review it here. The vertex operator \eqref{het} transforms as a primary field with conformal weight zero after using the equation of motion for the auxiliary fields. Finally, the  quadratic term  takes the form

\begin{equation}
\begin{aligned}
S =\frac{1}{4} \int d^{10}x & \left[  -\frac{1}{4}G^{(1)mn}\bx^{2} G^{(1)}_{mn} - \frac{1}{2} \bx(\pd_{r}G^{(1)nr})(\pd^{s}G^{(1)}_{sn})   - \frac{1}{5}\bx G^{m(1)}_{m}\pd^{m}\pd^{n}G^{(1)}_{mn}   +\right. \\
&\left.- \frac{2}{5}(\pd^{m}\pd^{n}G^{(1)}_{mn})^{2}+\frac{1}{10} G^{r(2)}_{r}  (-\bx G^{m(1)}_{m}     -2\pd^{m}\pd^{n}G^{(1)}_{mn}) -6B^{(2)mn}\pd^{p}C_{mnp}  +\right. \\
&\left. +6C^{mnp}(-\frac{1}{2}\bx C_{mnp} + \frac{1}{4}\pd_{[p}\pd^{r}C_{mn]r})+ 2A^{am}(\bx A^{a}_{m} - \pd_{m}(\pd \cdot A^{a}))  \right],
\end{aligned}
\label{17}
\end{equation}
\noindent
where $\pd_{[p}C_{mn]r} = 2\pd_{p}C_{mnr} + 2\pd_{m}C_{npr} + 2\pd_{n}C_{pmr} $.

To write the action \eqref{17} in terms of gauge invariant objects, we  redefine the fields

\begin{equation}
\begin{aligned}
& G^{(1)}_{mn} = h_{mn} -\frac{1}{4}\eta_{mn}h_p^p, \quad G^{r(2)}_{r} = t - \frac{1}{4}\bx h_p^p \quad \Rightarrow \quad\delta h_{mn} = -\frac{1}{2}\pd_{(m}\Lambda_{n)}, \quad \delta t =0 .
\end{aligned}	  
\end{equation}

Using the field strengths for the gauge and 2-form fields together with the linearized Riemann tensor 
\begin{equation}
\begin{aligned}
& R_{abcd} = \pd_{b}\pd_{c}h_{ad}+\pd_{a}\pd_{d}h_{bc}-\pd_{a}\pd_{c}h_{bd}-\pd_{b}\pd_{d}h_{ac},\\
& F^{a}_{mn} = \pd_{m}A^{a}_{n}-\pd_{n}A^{a}_{m},\\
& H_{mnp} =  \pd_{p}B^{(2)}_{mn} + \pd_{m}B^{(2)}_{np} + \pd_{n}B^{(2)}_{pm},
\end{aligned}
\end{equation}

\noindent
the action \eqref{17} takes the form

\begin{equation}
\begin{aligned}
S = -\frac{1}{4} \int d^{10} x &\left[ \frac{6}{10}R_{mn} R^{mn}+ \frac{1}{10}R_{mnpq}R^{mnpq} + \frac{1}{5}tR-2C^{mnp}H_{mnp}+\right.\\
&\left.  \qquad\qquad\qquad- 3C^{mnp}\left(\bx C_{mnp}  - \frac{1}{2}\pd_{[p}\pd^{r}C_{mn]r} \right) + F^{amn}F_{mn}^{a}\right].
\end{aligned}
\end{equation}

Although the heterotic ambitwistor action correctly describes Yang-Mills, it also has a symmetric two-form field $h_{mn}$ whose kinetic action is neither Einstein nor conformal gravity. In addition, it contains an antisymmetric 2-form $B^{(2)}_{mn}$ and antisymmetric 3-form $C_{mnp}$ with unusual couplings.
It is interesting to note, however, that the heterotic ambitwistor string was used in \cite{Azevedo:2017lkz} to reproduce $MHV$ amplitudes for conformal gravity in $D=4$.

\paragraph{Acknowledgments:}
NB and ML would like to thank Thales Azevedo, Henrique Flores, Max Guillen and Warren Siegel for useful discussions. ML would also like to thank FAPESP grant 2016/16824-0 for financial support and NB thanks FAPESP grants 2016/01343-7 and 2014/18634-9 and CNPq grant 300256/94-9 for
partial financial support.

\end{document}